\documentclass[aps,prl,twocolumn,superscriptaddress]{revtex4}
\usepackage{amssymb}
\usepackage{graphicx}
\usepackage{amsmath}
\usepackage{natbib}
\usepackage{bm}

\usepackage{color}

\begin{document}

\title{Superradiant Mott Transition}
\author{Yu Chen}
\affiliation{Department of Physics, Capital Normal University, Beijing, 100048, China}

\date{\today }

\begin{abstract}
The combination of strong correlation and emergent lattice can be achieved when quantum gases are confined in a superradiant Fabry-Perot cavity. In addition to the discoveries of exotic phases, such as density wave ordered Mott insulator and superfluid, a surprising kink structure is found in the slope of the cavity strength as a function of the pumping strength. In this Letter, we show that the appearance of such a kink is a manifestation of a liquid-gas like transition between two superfluids with different densities. The slopes in the immediate neighborhood of the kink become divergent at the liquid-gas critical points and display a critical scaling law with a critical exponent 1 in the quantum critical region. Our predictions could be tested in current experimental set-up.

\end{abstract}

\maketitle

\emph{Introduction}. 
The capability of confining quantum gases in a cavity enables us to realize strongly coherent coupling between atoms and light \cite{Esslinger07, Colombe07}. With this techniqual advance, the Dicke model for superradiance has finally been achieved experimentally in cold atom system \cite{Esslinger10} after decades of searching \cite{Dicke}. What goes beyond the physics of the original Dicke model in these experiments is that an emergent lattice appears in concurrence with the superradiance and the atoms self-orgarnize themselves into a density pattern. A roton mode softening across the superradiance transition is observed as a signiture of this emergent property\cite{Esslinger12}.
 
By imposing an optical lattice on an atomic gas, many models describing strongly correlated physics can be realized in a cold atomic setting, such as the Bose Hubbard model and Fermi Hubbard model \cite{OpticalLatticeReview}. A novel aspect of the emergent optical lattice, compared to an externally imposed one, is that a long-range interaction between the atoms could be established. As a result, when strong interaction meets emergent lattices, competitions between local interactions and long range interactions appear. Recently, this combination has been achieved by Hamburg and ETH's experimental groups by loading strongly interacting Bose gases into a strong coupling cavity\cite{Hemmerich15ex, Esslinger16}. Exotic phases like density ordered superfluid and Mott insulator are observed as a manifestation of the competition between local onsite interactions and cavity mediated long range interactions, as is predicted by quite a few theoretical works\cite{Simons09,Simons10,Ciuti13,He13,Hemmerich15th}.

Limited by technique, precise experimental knowledge of phase transitions between different phases still lacks. Very recently,  one pioneering work  explored  the phase boundaries and metastable states in the neighborhood of the phase transition between a homogenous Mott insulator and a density ordered Mott insulator\cite{Esslinger17}.
These metastable states are indictions of the first order transitions predicted by recent theoretical studies of phase boundaries based on ETH's set-up\cite{Yu16,Brennecke16,Mueller16}. 
Although present theoretical studies are satisfactory in many aspects, one striking feature in the Hamburg's experiment remains to be explained, that is the sharp kink structure (large slope change) of the superradiant cavity field against the pumping strength in the vicinity of SF-to-MI transition \cite{Hemmerich15ex}. In this Letter, we construct an effective field theory close to the SF-to-MI transition point in the superradiant phase and give an explanation for the presence of these kinks. Our results are summarized as follows: (1) There is a  liquid-gas like transition between two superfluids with density difference; (2) Sharp kinks are present in a large region around the critical point which ends at the liquid-gas like transition; (3) The kink strength is divergent at the critical point with a critical exponent as $1$.  Our prediction of a liquid-gas like transition can be tested in current experimental set-ups and the appearance of divergent kinks in superfluid phase serves as the smoking-gun.
\begin{figure}[t]
\includegraphics[width=6.6cm]{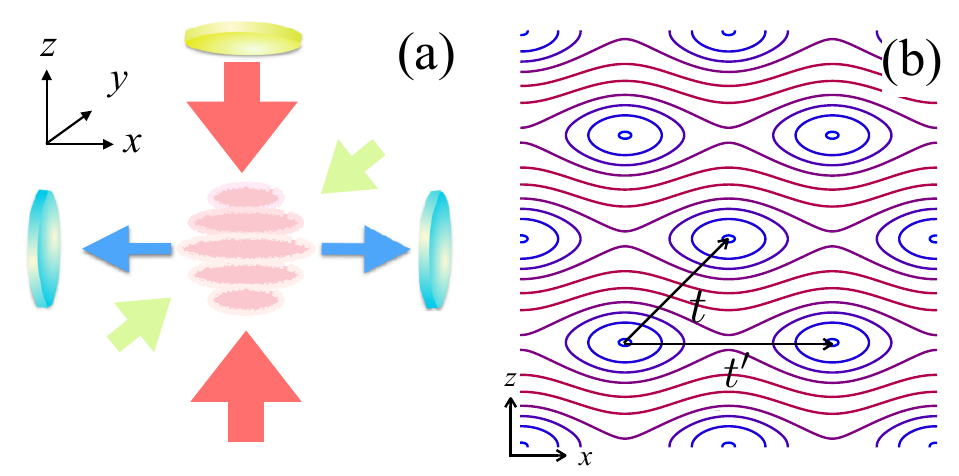}
\caption{(a) Scheme of single mode cavity with interacting Bose gases. The pumping field is in z direction, and cavity field is in x direction. (b) Emergent lattice configuration on xz plane in a superradiant cavity.}\label{Fig:Setup}
\end{figure}

{\emph {Model}.} 
We focus on the experiments by the Hamburg's group, where clear kink structures are observed. The experimental set up is shown in Fig.\ref{Fig:Setup} (a)\cite{Hemmerich15ex}: the pumping laser field is aligned in the $\hat{z}$ direction and linearly polarized in the $\hat{y}$ direction, while the cavity field is along the $\hat{x}$ direction and also polarized in the $\hat{y}$ direction. The cavity decay rate is $\kappa$. An additional off-resonant laser is applied in the $\hat{y}$ direction. The optical lattice the atoms experienced is $V_L({\bf r})=-V_P\cos^2 (k_0z)-V_y\cos^2 (k_0y)$ where $V_P$ is the pumping field strength, $V_y$ is the strength of the additional laser field and $k_0$ is the wave vector of laser field. The interaction between cavity field $\hat{a}$ and the atoms gives rise to an extra lattice $V_C({\bf r })=\eta\cos (k_0x)\cos (k_0z)(\hat{a}^\dag+\hat{a})+U_0\cos^2 (k_0x)\hat{a}^\dag\hat{a}$. As long as the atom-cavity interaction strength $|\eta|=\sqrt{-V_P U_0}$ is sufficiently large, the cavity field $\hat{a}$ will condense to a coherent superradiance state with $\alpha=\langle \hat{a}\rangle$, where $\langle\cdot\rangle$ denotes an average over the steady state. We note that in the present experiment no external optical lattice exists along the axial direction of the cavity, in contrast with the set-up in the ETH experiments. As a result, we expect that the feedback from the cavity field in the Hamburg set-up is much stronger than that in the ETH set-up.  

The optical lattice $V_L({\bf r})+V_C({\bf r})$ has a tipical potential contour in the $xz$ plane as shown in Fig.\ref{Fig:Setup} (b).
We project the motion of the atoms to the lowest band of $V_L({\bf r})+V_C({\bf r})$, and obtain a tight-binding model
\begin{eqnarray}
H&=&-t\sum_{\langle {\bf ij}\rangle_{xz}}(\hat{b}_{\bf i}^\dag \hat{b}^{}_{\bf j}+h.c.)-t_y\sum_{\langle {\bf i}{\bf j} \rangle_{y}}(\hat{b}_{\bf i}^\dag\hat{b}_{{\bf j}}^{}+h.c.)\nonumber\\
&&-t'\sum_{\langle\langle{\bf ij}\rangle\rangle_x}(\hat{b}_{\bf i}^\dag \hat{b}^{}_{\bf j}+h.c.)+H_{\rm MI}-\delta_c|\alpha|^2,\label{h}\\
H_{\rm MI}&=&\frac{U}{2}\sum_{\bf i}\hat{n}_{\bf i}(\hat{n}_{\bf i}-1)-\mu_\alpha\sum_{\bf i}\hat{n}_i
\end{eqnarray}
where the field operator ${b}_{\bf i}$ annihilates a boson on site ${\bf i}$ [$=(i_x,i_y,i_z)(\pi/k_0)$]. Here $\langle {\bf ij}\rangle$ and $\langle\langle{\bf ij}\rangle \rangle$ denote the nearest neighbor (NN) and the next nearest neighbor (NNN) respectively; the additional subindices $x,y,z$ restrict the neighboring sites along these specific directions. The parameters $t$ and $t_y$ are the NN hopping strength in the $xz$ plane and along the $y$-direction respectively, and $t'$ is the NNN hopping strength along the $x$-direction.  The chemical potential $\mu_\alpha=\mu-\eta(\alpha+\alpha^*)-U_0|\alpha|^2$ includes the onsite energy shift due to the cavity field and $U$ is the onsite interaction strength\cite{Note}. The total lattice site number is $N_\Lambda$.  

\emph{ Effective Theory}. To study phase transitions in the presence of the cavity field $\alpha$, we proceed to derive from Eq.~(\ref{h}) an effective field theory involving $\alpha$ and the low energy degrees of freedom of the bosons.

To begin, we introduce a local superfluid order parameter $\varphi \equiv W\langle b_{\bf i}\rangle$ where $W=4t+2t'+2t_y$. We note that due to the long range coherence, the superfluid phase $\varphi$ is site-independent. Assuming we are in the vicinity of SF-to-MI transition where superfluid order is weak,  we can obtain an effective mean field theory of $\varphi$ perturbatively from Eq.~(\ref{h}) as follows \cite{Sachdev}. We first diagonalize the on-site Hamiltonian $H_{\rm MI}=\frac{U}{2}\sum_{\bf i}n_{\bf i}(n_{\bf i}-1)-\mu_\alpha\sum_{\bf i}n_i$ to obtain the Mott eigenstates $|\ell\rangle$, where $\ell$ is the atomic site occupation number. Then, by approximating the tunnelling part in Eq.~(\ref{h}) at the mean field level by $b_{\bf i}^\dag b^{}_{\bf j}=(\varphi^*b^{}_{\bf j}+b_{\bf i}^\dag\varphi)/W-|\varphi|^2/W^2$, we calculate the energy correction to the Mott eigenstates due to such a tunneling term.

Here we focus on the energetically degenerate point of two adjacent Mott insulator phases with occupation number $\ell$ and $\ell+1$. At such a point, $\epsilon_{\ell}=\epsilon_{\ell+1}$ leading to $\mu_\alpha=\nu_\ell=U\ell$, where $\epsilon_{\ell}\equiv\langle\ell|\hat{H}_{\rm MI}|\ell\rangle/N_\Lambda$. To obtain accurate results around $\mu_\alpha\approx \nu_\ell$, we follow three steps. First, we carry out the non-degenerate perturbation in the subspace of $\{ |\ell\rangle, |\ell-1\rangle\}$ and $\{ |\ell+1\rangle, |\ell+2\rangle\}$ to find two ``restricted ground states" $|L\rangle$ and $|R\rangle$ in each subspace.
Second, we reconstruct a reduced hamiltonian in a Hilbert space spanned by $|L\rangle$ and $|R\rangle$. Accurate ground state energy could be obtained by diagonalizing this $2\times2$ matrix and the corresponding energy density is  
\begin{eqnarray}
\mathcal E&=&\!\!-\frac{{\delta}_c}{N_\Lambda}|\alpha|^2+(\epsilon_{\ell}+\epsilon_{\ell+1})/2+r|\varphi|^2\nonumber\\
&&\!\!-\sqrt{\left((\epsilon_{\ell+1}-\epsilon_{\ell})/2-\bar\chi|\varphi|^2\right)^2+(\ell+1)|\varphi|^2},\label{e2}
\end{eqnarray}
where $r=1/W-\ell/(\epsilon_{\ell-1}-\epsilon_{\ell})-(\ell+2)/(\epsilon_{\ell+2}-\epsilon_{\ell+1})$, $\bar{\chi}=(\ell+2)/(\epsilon_{\ell+2}-\epsilon_{\ell+1})-\ell/(\epsilon_{\ell-1}-\epsilon_{\ell})$. 
Finally we insert in Eq.~(\ref{e2}) the steady state solution of cavity field $\alpha$ determined by $i\partial_t\alpha=\partial \langle\hat H\rangle/\partial\alpha^*-i\kappa\alpha=0$. Under the condition that the decay rate $\kappa$ is large compared with atomic recoil energy $E_r\equiv k_0^2/2m$ where $m$ is the mass of the atom, the latter equation gives $\alpha=\eta N_\Lambda\langle\hat{n}\rangle/({\delta}_c+i\kappa)$. Here $\langle\hat{n}\rangle$ is the average occupation number per site. In solving for $\alpha$, we have used the fact that $U_0N_\Lambda\langle \hat{n}\rangle/\delta_c \ll 1$ and neglected the higher order terms.  In this approximation, self-consistent steady state solutions and energy density minimum are identical to each other.

An important point in our treatment is that we have eliminated the cavity strength $\alpha$ in favor of the occupation number $\langle n\rangle $ by means of the steady state equation. In doing so we can express the energy density as a function of superfluid order parameter $\varphi$ and the occupation number deviation 
\begin{eqnarray}
\theta\equiv\langle\hat{n}\rangle-({\ell}+\frac{1}{2}).
\end{eqnarray}
As we shall see shortly, the above choice of $\theta$ as the additional order parameter is useful in revealing certain hidden symmetry of the system. By replacing $\alpha$ with $\theta$, we get the energy density as
\begin{align}
\mathcal E&={\cal E}_{c}\theta^2+r|\varphi|^2\nonumber\\
&-\sqrt{(\ell+1)|\varphi|^2+({\cal E}_{c}\theta+U\delta+\bar{\chi}|\varphi|^2)^2}+{\cal O}(\varphi^4),\label{Eqn:EffectiveTheory}
\end{align}
where ${\cal E}_{c}\equiv-{\delta}_c N\eta^2/({\delta}_c^2+\kappa^2)>0$, and $\delta=(\mu+2{\cal E}_c(\ell+1/2)-U\ell)/2U$; the detuning $\delta$ quantifies the deviation from the degenerate point $\mu_\alpha=\nu_\ell$. Note that $r$ and $\bar\chi$ are implicit functions of $\theta$. Details of the derivation Eq.~(\ref{Eqn:EffectiveTheory}) are relegated to Ref\cite{Note}. 
Eqn(\ref{Eqn:EffectiveTheory}) is the effective theory for our following analysis of the superradiant Mott transition.

\emph{Liquid-gas like transition}. In the limit of large pumping field strengths, the optical lattice potential is deep and the hopping strengths are small such that $r$ is positive and large. As a result the energy density $\mathcal E$ is minimized at $\varphi=0$ for any $\theta$, namely the system is in the Mott insulator phase. In this case we have
\begin{align}
\mathcal E&={\cal E}_{c}\theta^2-|{\cal E}_{c}\theta+U\delta|.\label{MItoMI}
\end{align}
As expected, $\mathcal E$ is further minimized at $\theta=1/2$ ($\langle n\rangle=\ell+1$) for positive detuning $\delta$ and at $\theta=-1/2$ ($\langle n\rangle=\ell$) for negative $\delta$. Right at $\delta=0$, $\mathcal E$ is symmetric under the transformation $\theta\rightarrow -\theta$. We emphasize that this symmetry is an emergent symmetry and its appearance is responsible for the properties we shall describe below.
\begin{figure}[t]
\includegraphics[width=6.2cm]{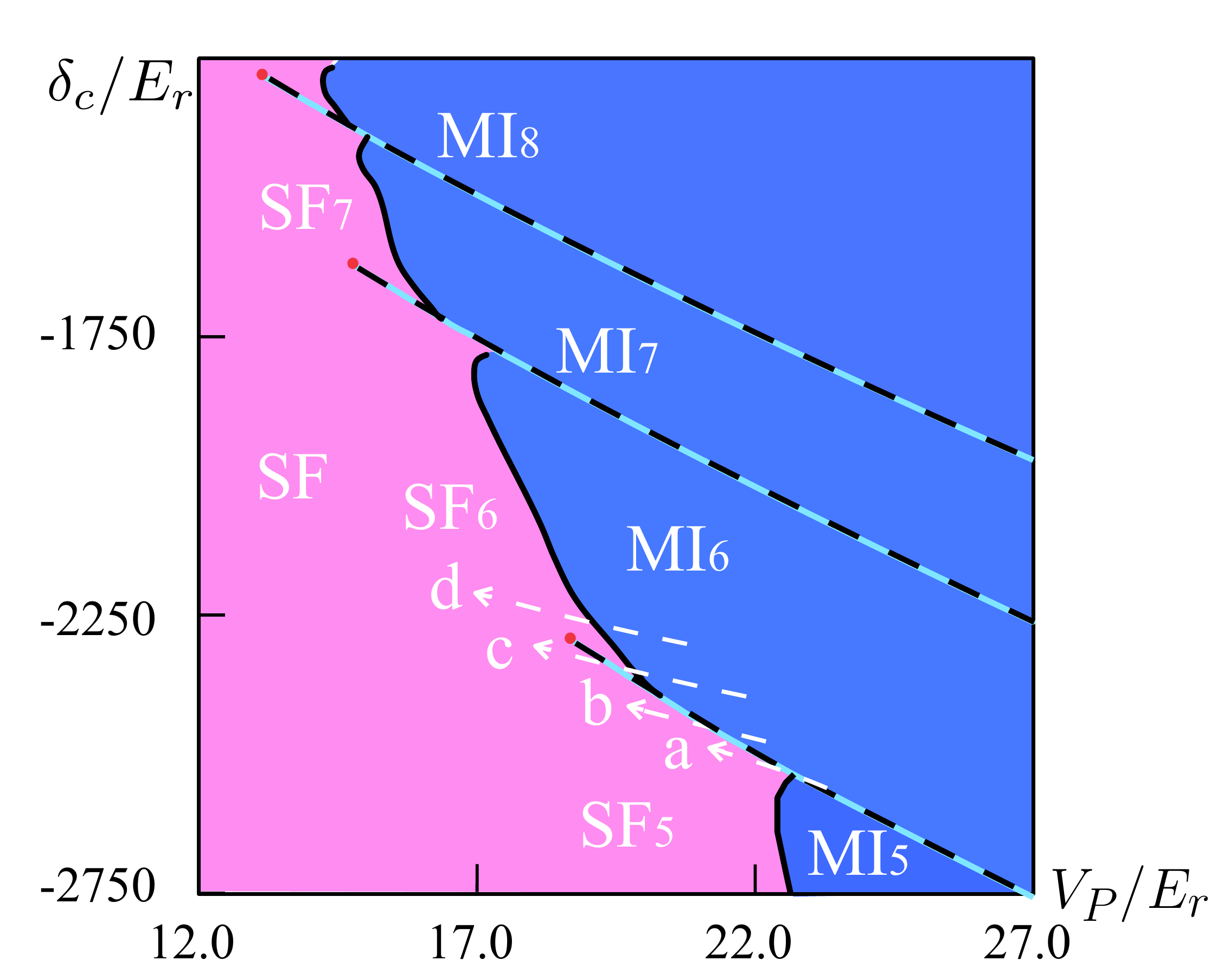}
\caption{Phase diagram of Mott transition in superradiance phase. SF phase is in red while MI phases are in blue. Black lines represent second order transitions and the blue black dashed lines represent first order transitions. ``Liquid-gas" critical point is marked in red dot. White dashed lines are the paths for $\delta=0, 0.005, 0.011$ and $0.0125$. ${\rm MI}_\ell$ labels Mott insulator phase with density $\ell$, ${\rm SF}_\ell$ labels superfluid phase with approximate density $\ell$. }
\label{PhaseDiagram}
\end{figure}
%

In superfluid phase where $\varphi\neq0$, the exact symmetry with respect to $\theta\rightarrow-\theta$ for $\delta=0$ no longer exists. However, by laws of continuity, an approximate ``$\theta$ reflection symmetry" remains if both $\varphi$ and $U\delta$ are small. For large $|U\delta|$, which is far away from the symmetric line, we carry out a Taylor expansion in terms of of $\varphi$ in Eqn.(\ref{Eqn:EffectiveTheory}) and get 
\begin{align}
\mathcal E=& {\cal E}_{c}\theta^2-|{\cal E}_{c}\theta+U\delta|+\nonumber\\
&\!\!\left(r-\frac12\frac{(\ell+1)+2(\mathcal E_{c}\theta+U\delta)\bar\chi}{|\mathcal E_{c}\theta+U\delta|}\right)|\varphi|^2+u_4|\varphi|^4,\label{Eqn:EffectSmallPhi}
\end{align}
where the $u_4|\varphi|^4$ term with $u_4>0$ is added phenomenlogically. For large and negative $U\delta$, the minimum of ${\cal E}$ is located on the $\theta>0$ side. A SF-to-MI transition is triggered when $r-(\ell+1)/2({\cal E}_c\theta+U\delta)-\bar{\chi}$ changes sign. On the other hand, for large and positive $\delta$, the minimum ${\cal E}$ is located at $\theta<0$ side, where a SF-to-MI transition occurs when $r+(\ell+1)/2({\cal E}_c\theta+U\delta)+\bar{\chi}$ changes sign. These two transitions are traditional second order Landau-Ginzberg transitions between supefluid and Mott insulator. The boundaries of the transitions are shown as black lines in Fig.\ref{PhaseDiagram}.
\begin{figure}[t]
\includegraphics[width=8.cm]{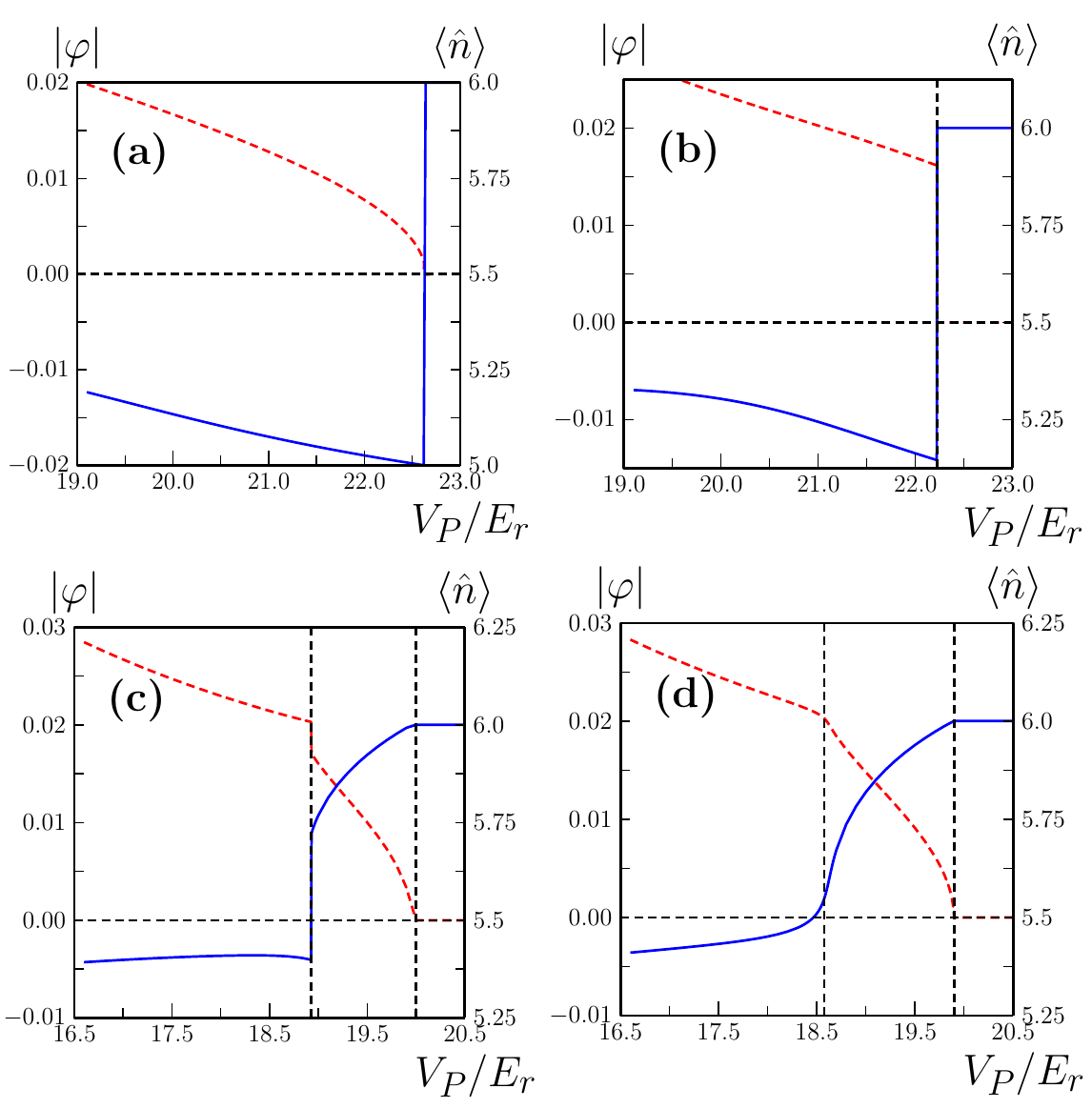}
\caption{ In (a), (b), (c), (d), the order parameters $\varphi$(dashed red) and $\langle n\rangle$(solid blue) along paths $\delta=0, 0.005, 0.011$ and $0.0125$ (labeled as a, b , c, d in Fig.\ref{PhaseDiagram} as white dashed lines) are given. We use black dashed lines to label phase transition points. In (a), the SF-to-MI transition happens when $\varphi$ becomes nonzero; In (b), a first order transition is labeled; In (c), there are two transitions; In (d), the right side dashed line is a SF-to-MI transition and the left side dashed line labels the kink.}
\label{FreeEn}
\end{figure}

In the other limit, namely for $\delta\approx0$, we are in a highly symmetric region with an emergent ${\mathbb Z}_2$ symmetry. In this region we assume that $|\varphi|$ is relatively large and ${\cal E}_c\Theta\equiv{\cal E}_c\theta+U\delta+\bar{\chi}|\varphi|^2$ is relatively small. We then carry out a Taylor expansion of $\Theta$ in Eqn(\ref{Eqn:EffectiveTheory}) and get
\begin{eqnarray}
{\cal E}&=&{\cal E}_0\!-\!2(U\delta\!+\!\bar{\chi}|\varphi|^2)\Theta\nonumber\\
&&\!+\!({\cal E}_c\!-\!\frac{{\cal E}_c^2}{2\sqrt{\ell\!+\!1}|\varphi|})\Theta^{2}+\frac{{\cal E}_c^4}{4\sqrt{(\ell\!+\!1)^3}|\varphi|^3}\Theta^4,
\end{eqnarray}
where ${\cal E}_0=r|\varphi|^2\!-\!\sqrt{\ell\!+\!1}|\varphi|\!-\!\frac{(U\delta\!+\!\bar{\chi}|\varphi|^2)^2}{{\cal E}_c}\!+\!u_4|\varphi|^4$.   At a fine-tuned point where the linear term in the above expression disappears and the ``$\theta$ reflection symmetry" is restored, a second order transition is triggered by ${\cal E}_c-{\cal E}_c^2/\sqrt{\ell+1}|\varphi|\leq0$. In this scenario, the symmetry is spontaneously broken, leading to either a positive $\Theta$ for ``liquid" superfluid with high density and a negative $\Theta$ for the ``gas" superfluid with low density. A first order transition between two SFs could be engendered by the sign change of the linear $\Theta$ term, in analogy with the addition of an external magnetic field in a ferromagnet. A first order transition between a high density Mott insulator to low density superfluid is also possible as a continuation of liquid-gas like transition between two SFs. The existence of the liquid-gas like transition is the consequence of the emergent ${\mathbb Z}_2$ ``$\theta$ reflection symmetry" being broken.

Additional numerical simulation can be carried out based on our effective field theory up to $|\varphi|^4$ order. The phase diagram is shown in Fig.\ref{PhaseDiagram} where black and the black blue dashed lines represent the boundaries of the second order and the first order transitions, respectively; the red points represent the critical points. Four routes a, b, c and d for small $\delta$ around ``$\theta$ reflection symmetric" region are taken across these boundaries and the order parameters along them are shown in Fig.\ref{FreeEn}. Along route a, $\delta=0$, a second order SF-to-MI transition is trigger by lowering the puming strength. For $\delta=0$ and large pumping strength, the system is on the $\theta$ reflection symmetric line where two Mott insulators are degenerate. Along route b, a first order transition between Mott insulator and superfluid is displayed by a jump of the order parameter  in both $\varphi$ and $\langle \hat{n}\rangle$. Along route c, there are two transitions. From right to left, the first one is a second order SF-to-MI transition by a spontaneous breaking of the U(1) symmetry in $\varphi$; the second one is the liquid-gas like transition between two SFs. Finally, along path d, there is only a second order MI-to-SF transition. But there is an obvious kink in both superfluid order and density order.
\begin{figure}[t]
\includegraphics[width=7.8cm]{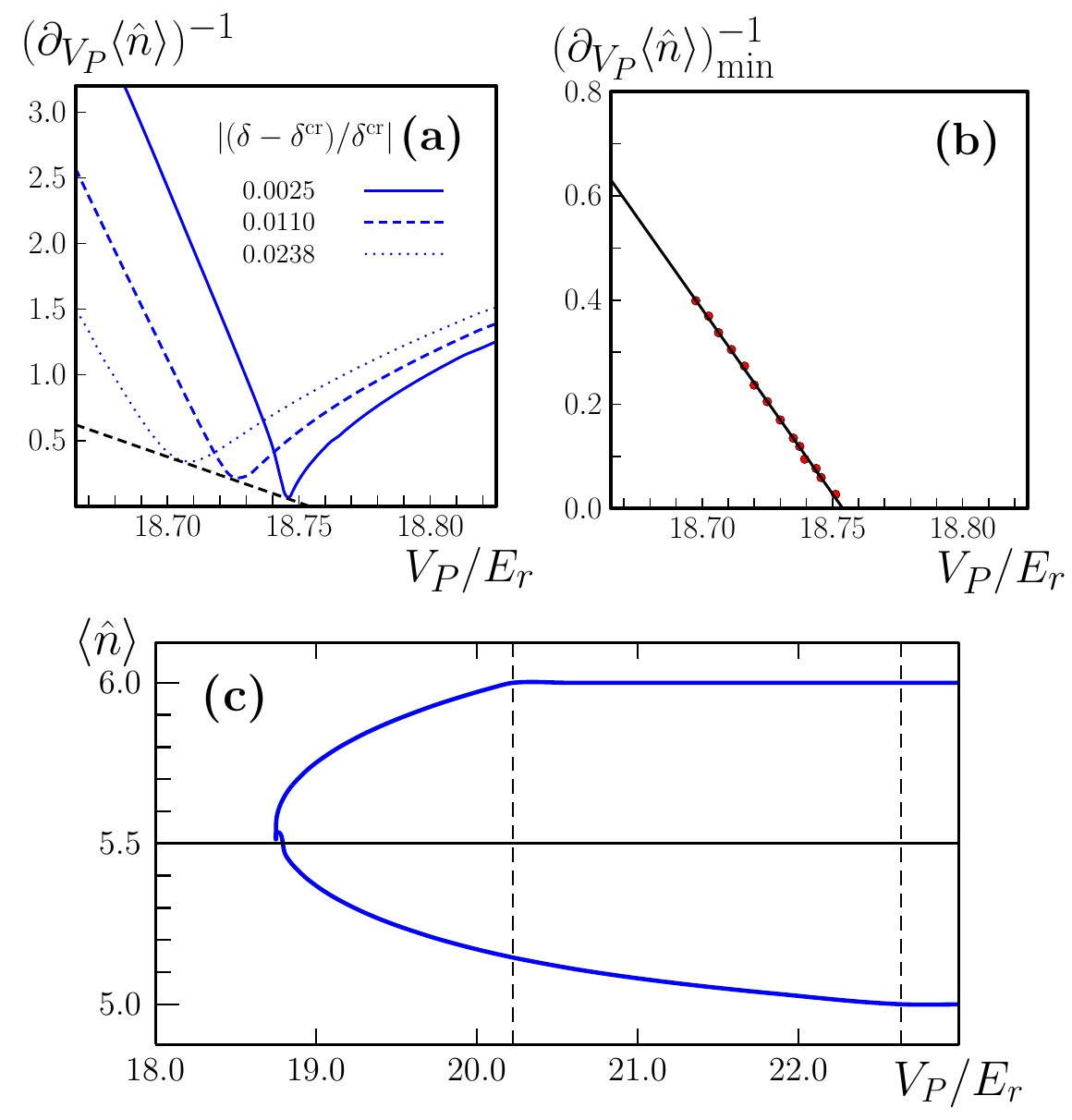}
\caption{(a) $(\partial_{V_P}\langle \hat{n}\rangle)^{-1}$ as a function of pumping field strength $V_P$ for different $\delta$. $\delta^{\rm cr}$ is the imbalance $\delta$ at critical point. (b) The minimal of $(\partial_{V_P}\langle \hat{n}\rangle)^{-1}$ for different $\delta$ as a function of $V_P$. The correlation coefficient of linear scaling is 0.9997. (c) The density of two phases on each side of the transition line as a function of pumping field strength. The density difference plays a role of order parameter for liquid-gas like transition. One can also see the critical point for divergent kink and starting of ``density splitting order" are identical.  }
\label{Kink}
\end{figure}

Here we identify the kink in the derivative of density order against pumping strength $\partial_{V_P}\langle n\rangle$. We find a critical scaling law as $|V_P-V_{P}^{\rm cr}|^{-1}$ for the maximal $\partial_{V_P}\langle n\rangle$, which is divergent at critical point. Here $V_P^{\rm cr}$ is the pumping strength for the critical point. This criticality is shown in Fig.\ref{Kink} (a) and (b). Here we could observe that $\partial\langle n\rangle/\partial\delta$ shares a similarity with the magnetic susceptibility in magnetic systems and has a physical meaning of compressibility. Therefore the $|V_P-V_P^{\rm cr}|^{-1}$ scaling law is a reflection of universal critical exponent $\gamma$ for compressibility.

Concerning experimental observation, the negative kink is more difficult to observe for technical reasons. As is shown by the white dashed line d in Fig.\ref{PhaseDiagram} for fixed $\delta$, the route we take always bypasses the critical point from above, where only positive kinks are accessed. This is consistent with the experimental procedure where similar routes are taken.

\emph{Conclusion and Outlook}. 
To conclude, we construct an effective field theory from microscopic model to study the Mott transition in emergent lattices and find a liquid-gas like transition between two superfluids. The liquid-gas like transition ends at a  critical point within superfluid phase, and a divergent density kink is predicted (so is the superradiance kink). This kink exists in a large region around the critical point and the maximal density slope scales as $|V_P-V_{P}^{\rm cr}|^{-1}$. We point out that in three dimensional system, $\gamma\approx1.24$ because the renormalization group flow drives the system away from Gaussian fixed point to Wilson-Fisher fixed point. Further, because of the strong quantum fluctuations in superfluid order $\varphi$ and the non-equilibrium nature of this system, there are extra corrections for $\gamma$.
This problem will be explored in a future study. Further experiments with the current set-up can be carried out to test the details of our predictions.

\textit{Acknowledgement.} We acknowledge Zhenhua Yu, Hui Zhai, and Andreas Hemmerich for inspiring discussions. We would like to thank Zhigang Wu and Ren Zhang for careful reading our manuscript and advises for presentation.  This work is supported by NSFC Grant No. 11604225 and Foundation of Beijing Education Committees under Grants No. KM201710028004.

\appendix
\begin{widetext}

\centerline{\bf Supplementary Material}
\section{Tight-binding parameters}
Considering the Pumping strength being $V_P$, and the condensed cavity field strength being $\alpha$. Then we optical lattice atom gas experienced at $y=0$ plane is
\begin{eqnarray}
V_P({\bf r})+\eta({\bf r})(\alpha+\alpha^*)\cos k_0x\cos k_0z+U_0({\bf r})|\alpha|^2
\end{eqnarray}
where $V_P({\bf r})=-V_P\cos^2 k_0x$, $\eta({\bf r})=\eta\cos k_0x\cos k_0z$, $U_0({\bf r})=U_0\cos^2 k_0z$.
Let us suppose $|\alpha|$ is large. One can see in even lattice, around $(0,0)$ point, the potential in x direction is $V_P\cos^2k_0x+\eta(\alpha+\alpha^*)\cos k_0x+U_0\alpha^*\alpha$, the potential in z direction is $-V_P+\eta(\alpha+\alpha^*)\cos k_0z+U_0\alpha^*\alpha \cos^2 k_0z$. Around $(\pi/k_0,0)$ point, the potential along x direction is $-V_P\cos^2k_0x+\eta(\alpha+\alpha^*)\cos k_0x+U_0\alpha^*\alpha$, while the potential along z direction is $-V_P-\eta(\alpha+\alpha^*)\cos k_0z+U_0\alpha^*\alpha\cos^2 k_0z$. When $\eta(\alpha+\alpha^*)-2U_0\alpha^*\alpha<0$ or $-2V_P+\eta(\alpha+\alpha^*)<0$, the odd lattice can not host any particles. As $\eta=-\sqrt{-U_0V_P}$, the condition is $\sqrt{-V_p/U_0}>2\alpha^*\alpha/(\alpha+\alpha^*)$ or $\sqrt{-V_P/U_0}<(\alpha+\alpha^*)/2$. If the cavity decay rate $\kappa$ is zero, these two conditions are contradictory to each other. Therefore for small $\kappa$ (large compared to recoil energy, small compare to cavity detune), the emergent lattice could only host particle on even site.

Now suppose the parameters are in one site region, then two hopping strength is vital, one from $(0,0)$ to $(\pm\pi/k_0,\pm\pi/k_0)$, another is from $(0,0)$ to $(0,\pm2\pi/k_0)$. For the first case, we take the route $(x,x)$, then the potential becomes $(V_P+\eta(\alpha+\alpha^*)+U_0\alpha^*\alpha)\cos^2 k_0x$. In WKB approximation, the hopping strength calculated by instanton tunneling, which is
\begin{equation}
t=E_r\left(|-V_P+2\eta{\rm R}\alpha+U_0|\alpha|^2|/2E_r\right)^{3/4}e^{-S_0},
\end{equation} 
where $S_0=\int_0^{\pi/k_0}dx\sqrt{2m|-V_P+2\eta{\rm R}\alpha+U_0|\alpha|^2|}\sin k_0x=2\sqrt{|-V_P+2\eta{\rm R}\alpha+U_0|\alpha|^2|/E_r}$. Along $x=0$, the potential is $-V_P+\eta(\alpha+\alpha^*)\cos k_0z+U_0\alpha^*\alpha\cos^2 k_0z$, correspondingly, we get
\begin{eqnarray}
t'&=&E_r\left(\frac{4|\eta|{\rm R}\alpha}{E_r}\right)^{3/4}e^{-S'_0},
\end{eqnarray}
where
$S'_0=\int_0^{2\pi/k_0}dz \sqrt{2m(2|\eta|{\rm R}\alpha(1-\cos k_0z)+|U_0||\alpha|^2\sin^2 k_0z)}=2 \sqrt{\frac{4|\eta|{\rm R}\alpha}{E_r}}\left(\sqrt{1+\frac{U_0|\alpha|^2}{\eta{\rm R}\alpha}}+\frac{{\rm Arcsinh}\sqrt{U_0|\alpha|^2/\eta{\rm R}\alpha}}{\sqrt{U_0|\alpha|^2/\eta{\rm R}\alpha}}\right)\approx 4\sqrt{\frac{4|\eta|{\rm R}\alpha}{E_r}}$.

For onsite interaction energy, we could approximate the Wannier wave function on even site by ground state wave function of harmonic trap. The harmonic trap on the even site is $\frac{1}{2}(\omega_x^2 x^2+\omega_y^2 y^2+\omega_z^2 z^2)$, $\omega_x=\sqrt{|-V_P+\eta{\rm R}\alpha|}$, $\omega_z=\sqrt{|\eta{\rm R}\alpha+U_0|\alpha|^2|}$, $\omega_y=\sqrt{|V_y|}$. Then the Wannier wave function is $\Phi({\bf r})=\phi_x(x)\phi_y(y)\phi_z(z)$, where $\phi_\zeta(\zeta)=(\omega_\zeta/\pi \hbar)^{1/4}e^{-\omega_\zeta \zeta^2/2\hbar}$. The interaction energy density is $g\int d^3{\bf r}\Phi^4({\bf r})\propto \sqrt{\omega_x\omega_y\omega_z}$. As the $\alpha$ dependence is only in $\omega_x$ and $\omega_z$, therefore we could get
\begin{eqnarray}
U=\bar{U} \sqrt{\omega_x\omega_z},
\end{eqnarray}
where interaction strength $\bar{U}$ could be fixed by tune scattering length between atoms and the depth of lattice in y direction. Compared with the hopping strength, the interaction strength density is just a slow varying function of $\alpha$. With all these parameters, we could get the tight binding model given in Eqn. (1).

\section{The equivalence of free energy minimal and steady state equation in large $\delta_c/\kappa$ limit}

In this section we are going to show that the steady state equation together with mean field equation for SF order could ensure the steady state solution minimize the free energy of the system in small $\kappa$ limit.
\begin{eqnarray}
i\partial_t\alpha=-\delta_c\alpha-i\kappa\alpha+\frac{\delta H}{\delta \alpha^*}=0\hspace{3ex}\frac{\partial H}{\partial \varphi^*}=\frac{\partial H}{\partial\varphi}=0
\end{eqnarray}

\begin{eqnarray}
\frac{d H}{d\varphi}\!\!&=&\!\!\frac{\partial H}{\partial\varphi}\!+\!\frac{\partial H}{\partial\alpha}\frac{d\alpha}{d\varphi}+\frac{\partial H}{\partial\alpha^*}\frac{d\alpha^*}{d\varphi}-\delta_c\frac{d |\alpha|^2}{d\varphi}\nonumber\\
\!\!&=&\!\!\frac{\partial H}{\partial\varphi}\!+\!(\delta_c-i\kappa)\alpha^*\frac{d\alpha}{d\varphi}\!+\!(\delta_c+i\kappa)\alpha\frac{d\alpha^*}{d \varphi}\!-\!\delta_c\frac{d|\alpha|^2}{d\varphi}=-i\kappa \alpha^*\frac{\partial\alpha}{\partial\varphi}+i\kappa \alpha\frac{\partial\alpha^*}{\partial \varphi}
\end{eqnarray}
Since in general we could write $\alpha=\Xi(\varphi)/(\tilde{\delta}_c(\varphi)+i\kappa)$. One could find that $dH/d\varphi=-\kappa^2/(\tilde{\delta}^2_c+\kappa^2)|\alpha|^2(d\tilde{\delta}_c/d\varphi)$. It is zero if $\kappa=0$ or $d\tilde{\delta}_c/d\varphi=0$. The former is zero dissipation limit. The latter means the phase of superradiance being locked by dissipation. When $|\delta_c/\kappa|\gg1$, the derivative $d\tilde{\delta}_c/d\varphi$ is relatively small and the global factor is suppressed. Therefore in this limit, steady state solutions and energy density minimum solutions are approximately equivalent.

\section{Deduction of Effective theory (upto $|\varphi|^4$ order)}
In this section, we give a deduction for effective theory up to $|\varphi|^4$ order. The effective theory with $|\varphi|^2$ term could be get by dropping all $|\varphi|^4$ terms. In the first step, we select two subspaces \{$|\ell\rangle$, $|\ell-1\rangle$, $|\ell-2\rangle$\} and \{$|\ell+1\rangle$, $|\ell+2\rangle$, $|\ell+3\rangle$\}. (Keep in mind that $|\ell\rangle$ and $|\ell+1\rangle$ are nearly degenerate) By non-degenerate perturbation within subspace $|\ell\rangle$, $|\ell-1\rangle$, $|\ell-2\rangle$ or in subspace $|\ell+1\rangle$, $|\ell+2\rangle$, $|\ell+3\rangle$, we get two ``ground states" in two restricted Hilbert spaces. 
\begin{eqnarray}
|L\rangle\!\!&=&\!\!{\cal N}_L\!\left(|\ell\rangle-\left(\frac{\sqrt{\ell}\varphi^*}{\epsilon_{\ell-1}-\epsilon_\ell}-\frac{\sqrt{\ell^3}|\varphi|^2\varphi^*}{(\epsilon_{\ell-1}-\epsilon_\ell)^3}+\frac{\sqrt{\ell}(\ell-1)|\varphi|^2\varphi^*}{(\epsilon_{\ell-1}-\epsilon_\ell)^2(\epsilon_{\ell-2}-\epsilon_\ell)}\right)|\ell-1\rangle\!+\!\frac{\sqrt{\ell(\ell-1)}(\varphi^*)^2}{(\epsilon_{\ell-1}-\epsilon_{\ell})(\epsilon_{\ell-2}-\epsilon_\ell)}|\ell-2\rangle\right)\\
|R\rangle\!\!&=&\!\!{\cal N}_R\!\left(|\ell+1\rangle-\frac{\sqrt{\ell+2}\varphi}{\epsilon_{\ell+2}-\epsilon_{\ell+1}}\!\left(\!1-\frac{(\ell+2)|\varphi|^2}{(\epsilon_{\ell+2}-\epsilon_{\ell+1})^2}+\frac{(\ell+3)|\varphi|^2}{(\epsilon_{\ell+3}-\epsilon_{\ell+1})^2}\!\right)\!|\ell+2\rangle\!+\!\frac{\sqrt{(\ell+2)(\ell+3)}(\varphi)^2}{(\epsilon_{\ell+2}\!-\!\epsilon_{\ell+\!1})(\epsilon_{\ell+3}\!-\!\epsilon_{\ell+\!1})}|\ell\!+\!3\rangle\!\!\right)
\end{eqnarray}
${\cal N}_L$ and ${\cal N}_R$ are normalizer ensuring $\langle L|L\rangle=\langle R|R\rangle=1$.

In the second step, we write down the reduced hamiltonian in the Hilbert space spanned by $|L\rangle$ and $|R\rangle$, which is
\begin{eqnarray}
\hat{h}_{\rm red}=-\frac{\delta_c}{N_\Lambda}|\alpha|^2+\frac{|\varphi|^2}{W}+\left(
\begin{tabular}{cc}
$\epsilon_\ell-\chi_L|\varphi|^2+u_L|\varphi|^4$&${\cal N}_L{\cal N}_R\sqrt{\ell+1}\varphi^*$\\
${\cal N}_L{\cal N}_R\sqrt{\ell+1}\varphi$&$\epsilon_{\ell+1}-\chi_R|\varphi|^2+u_R|\varphi|^4$
\end{tabular}
\right)
\end{eqnarray}
where $\chi_L=\ell/(\epsilon_{\ell-1}-\epsilon_{\ell})$,  $\chi_R=(\ell+2)/(\epsilon_{\ell+2}-\epsilon_{\ell+1})$,
\begin{eqnarray}
u_L&=&\frac{\ell^2}{(\epsilon_{\ell-1}-\epsilon_\ell)^3}-\frac{\ell(\ell-1)}{(\epsilon_{\ell-1}-\epsilon_\ell)^2(\epsilon_{\ell-2}-\epsilon_\ell)}\\
u_R&=&\frac{(\ell+2)^2}{(\epsilon_{\ell+2}-\epsilon_{\ell+1})^3}-\frac{(\ell+2)(\ell+3)}{(\epsilon_{\ell+2}-\epsilon_{\ell+1})^2(\epsilon_{\ell+3}-\epsilon_{\ell+1})}
\end{eqnarray}
Take the smaller eigen value of $\hat{h}_{\rm red}$, which gives us ground state energy density as
\begin{eqnarray}
{\cal E}=-\frac{\delta_c}{N_\Lambda}|\alpha|^2+\frac{\epsilon_\ell+\epsilon_{\ell+1}}{2}+r|\varphi|^2+u|\varphi|^4-\sqrt{{\cal N}^2_L{\cal N}^2_R(\ell+1)|\varphi|^2+\left(\frac{\epsilon_{\ell+1}-\epsilon_\ell}{2}-\bar{\chi}|\varphi|^2+\bar{u}|\varphi|^4\right)}\label{EnergyDensity}
\end{eqnarray}
where ${\cal N}_L^2{\cal N}_R^2=1-\ell|\varphi|^2/(\epsilon_{\ell-1}-\epsilon_\ell)^2-(\ell+2)|\varphi|^2/(\epsilon_{\ell+2}-\epsilon_{\ell+1})^2$ is approximated up to $|\varphi|^2$ order to make sure $|\varphi|^4$ order being exact. $r=1/W-(\chi_L+\chi_R)/2$, $\bar{\chi}=(\chi_R-\chi_L)/2$, $u=(u_L+u_R)/2$ and $\bar{u}=(u_R-u_L)/2$.

In the third step, we use steady state solution of  cavity field $\alpha=2\eta N_\Lambda(\ell+\frac{1}{2}+\theta)/(\delta_c+i\kappa)$ to rewrite the energy density as a function of $\theta$(Here we neglected  contributions from $t$ and $U$, because their effect on $\alpha$ is minor. We also use $\delta_c$ to replace $\tilde{\delta}_c$ to drop the frequency shift form $U_0$ term. As $|U_0N_\Lambda|\ll|\delta_c|$, $\tilde{\delta}_c\approx\delta_c$ stands, especially in large $|\delta_c|$ limit), one can get
\begin{eqnarray}
\frac{\epsilon_{\ell}+\epsilon_{\ell+1}}{2}-\frac{\delta_c}{N_\Lambda}|\alpha|^2&=&\!\!-\mu_{\ell+\frac{1}{2}}(\ell+\frac{1}{2})+\frac{U}{2}\ell^2+{\cal E}_c(\ell+\frac{1}{2})^2-{\cal E}_{c}\theta(2\ell+1)+{\cal E}_{c}\theta(2\ell+1)+{\cal E}_c\theta^2={\cal E}_0+{\cal E}_c\theta^2,\label{f1}
\end{eqnarray}
where ${\cal E}_c\equiv-\delta_c N_\Lambda\eta^2/(\delta_c^2+\kappa^2)$ and $\mu_{\ell+1/2}\equiv\mu+{\cal E}_c(2\ell+1)$. In a similar way we can get
\begin{eqnarray}
\frac{\epsilon_{\ell+1}-\epsilon_\ell}{2}&=&\frac{-(\mu-\eta(\alpha+\alpha^*))+U\ell}{2}=\frac{-\mu_{\ell+\frac{1}{2}}+U\ell-2{\cal E}_{c}\theta}{2}
=-{\cal E}_{c}\theta-U\delta,\label{f2}
\end{eqnarray}
where we define $2U\delta=\mu_{\ell+\frac{1}{2}}-U\ell$. 

By inserting Eqn.(\ref{f1}) and Eqn.(\ref{f2}) to Eqn.(\ref{EnergyDensity}) and change all $\alpha$ dependence in parameters to $\theta$ dependence, we get
\begin{eqnarray}
{\cal E}={\cal E}_c\theta^2+r|\varphi|^2+u|\varphi|^4-\sqrt{{\cal N}_L^2{\cal N}_R^2(\ell+1)|\varphi|^2+({\cal E}_c\theta+U\delta+\bar{\chi}|\varphi|^2-\bar{u}|\varphi|^4)^2}
\end{eqnarray}
By dropping $|\varphi|^4$ terms we get Eqn.(\ref{Eqn:EffectiveTheory}). Of course, Eqn.(\ref{Eqn:EffectiveTheory}) could also be get by the same method with smaller truncated Hilbert space.
\end{widetext}

\end{document}